\documentclass[prc,aps,preprint,showpacs,nofootinbib,showkeys,eqsecnum,tightenlines]{revtex4}    
\usepackage{epsfig}
\def\be{\begin{equation}}
\def\ee{\end{equation}}
\def\bea{\begin{eqnarray}}
\def\eea{\end{eqnarray}}
\newcommand{\noi}{\noindent}

\def\Journal#1#2#3#4{{#1} {\bf #2}, #3 (#4)}
\def\NPA{{ Nucl. Phys.} \bf A}
\def\NPB{{ Nucl. Phys.} \bf B}
\def\PRL{ Phys. Rev. Lett.\,\,}
\def\PRC{{ Phys. Rev.} \bf C}
\def\PRD{{ Phys. Rev.} \bf D}

\def\PLB{{ Phys. Lett.} \bf B}

\def\EPJA{{ Eur. Phys. J.}\bf A}

\def\IJMPE{{ Int. J. Mod. Phys.} \bf E}

\def\RMP{ Rev. Mod. Phys.\,\,}

\def\AJ{ Astrophys. J.\,\,}

\begin{document}
%
%
\title{
{\large{\bf Model dependence of the neutrino-deuteron
disintegration cross sections at low energies}}}
\author{B. Mosconi}
\affiliation{Universit$\grave{a}$ di Firenze, Department of
Physics, and Istituto Nazionale di Fisica Nucleare, Sezione di
Firenze, I-50019, Sesto Fiorentino (Firenze), Italy }
\author{P. Ricci}
\affiliation{Istituto Nazionale di Fisica Nucleare, Sezione di
Firenze, I-50019, Sesto Fiorentino (Firenze), Italy }
\author{E.~Truhl\'{\i}k}
\affiliation{Institute of Nuclear Physics ASCR, CZ--250 68
\v{R}e\v{z}, Czech Republic, and Institute for Nuclear Theory and
Department of Physics, University of Washington, Seattle, WA
98195, USA  }
\author{P. Vogel}
\affiliation{Kellog Radiation Laboratory and Physics Department,
Caltech, Pasadena, California, 91125, and Institute for Nuclear
Theory and Department of Physics, University of Washington,
Seattle, WA 98195, USA }

\begin{abstract}

Model dependence of the reaction rates for the weak breakup of
deuterons by low energy neutrinos is studied starting from the
cross sections derived from potential models and also from
pionless effective field theory. Choosing the spread of the
reaction yields, caused basically by the different ways the
two-body currents are treated, as a measure of the model dependent
uncertainty, we conclude  that the breakup reactions are $\sim$ 2
- 3 \% uncertain, and that  even the ratio of the charged to
neutral current reaction rates is also $\sim$ 2 \% uncertain.

\end{abstract}

\noi \pacs{  25.30.Pt, 11.40.Ha,  25.10.+s,   26.65.+t}

\noi \hskip 1.9cm \keywords{electron neutrinos; weak breakup;
deuteron; low energy}

\maketitle

\section{Introduction}
\label{intro}

The SNO collaboration \cite{SNO1,SNO2,SNO3,SNO4}, following the
original suggestion by late Herb Chen \cite{Chen85}, has
convincingly shown that the flavor of solar neutrinos is not
conserved. This was achieved by determining the yield of the
deuteron disintegration in both neutral and charged current
channels: \bea
\nu_x\,+\,d\,&\longrightarrow&\,\nu'_x\,+\,n\,+\,p\,, \label{NCN} \\
\nu_e\,+\,d\,&\longrightarrow&\,e^-\,+\,p\,+\,p\,.  \label{CCN}
\eea The neutrino flux deduced from the neutral current reaction
(\ref{NCN}) agrees within errors with the Standard Solar Model
(SSM) \cite{BP}, while the flux deduced from the charged current
reaction (\ref{CCN}) is smaller than the SSM prediction by a
factor of $\sim$ 3. The only reasonable way to interpret this
result, and the other observations of solar neutrinos
\cite{HM,SK,K,SAGE,GALLEX,GNO}, is in terms of neutrino
oscillations. This conclusion becomes inescapable when the reactor
neutrino experiment KamLAND \cite{KLAND1,KLAND2} is included in
the corresponding fit.

In order to relate the yield of the reactions observed in SNO to
the corresponding solar neutrino flux one needs to know the
neutrino-deuteron breakup cross section. Consequently, the cross
sections of the reactions (\ref{NCN}), (\ref{CCN}) and the
analogous ones initiated by antineutrinos, \bea
\overline{\nu}_x\,+\,d\,&\longrightarrow&\,\overline{\nu}'_x\,+\,n\,+\,p\,,
\label{NCA} \\
\overline{\nu}_e\,+\,d\,&\longrightarrow&\,e^+\,+\,n\,+\,n\,,
\label{CCA} \eea have been carefully evaluated during the past two
decades (see Refs.\,\cite{YHH,NSGK,NSAPMGK,MRT1,BCK,ASPFK} and
references therein). Here we wish to assess the uncertainties or
model dependence involved in these evaluations related to the
different ways the two-body exchange currents are treated.

The studies of the reactions (\ref{NCN})--(\ref{CCA}) at low
energies were performed in Refs.\,\cite{YHH,NSGK,NSAPMGK,MRT1}
based on the currents derived from elementary hadron amplitudes
extracted in the tree approximation from the chiral Lagrangians,
and using nuclear wave functions generated from realistic nuclear
potentials.

Alternatively, in Ref.\,\cite{BCK}, the cross sections derived in
the next-to-next-to-leading order of the pionless effective field
theory, were written in the form \be
\sigma_{EFT}(E_\nu)\,=\,a(E_\nu)\,+\,L_{1,A}\,b(E_\nu)\,.
\label{SEFF} \ee Tables of numerical values of the amplitudes
$a(E_\nu)$ and $b(E_\nu)$ are given in Ref.\,\cite{BCK} up to 20 MeV
in 1-MeV steps.

In principle, the effective field theory provides a more
fundamental approach to the study of nuclear phenomena, but it
contains parameters that cannot be determined in reactions between
elementary particles. The factor $L_{1,A}$ in Eq.\,(\ref{SEFF})
that parameterizes the effect of the isovector axial two-body
current, is an example of such a constant. Its value can be
determined from a measurement of any of the breakup processes
(\ref{NCN})--(\ref{CCA}). The analysis of various data
\cite{BY,BCV} provides $L_{1,A}$ value, however, with a large
error, \be L_{1,A}\,=\,3.6\,\pm\,5.5\,\,\,{\rm fm^3}\,.
\label{L1A} \ee

Alternatively, the value of $L_{1,A}$ can be determined by
comparing the cross sections  (\ref{SEFF}) with the cross sections
calculated employing the nuclear wave functions generated from
realistic one-boson-exchange-potentials (OBEPs) and the one- and
two-nucleon currents as it was done in the recent work
\cite{MRT1}. The resulting values of $L_{1,A}$ were confined
between the limits (see Table 2 in Ref.\,\cite{MRT1}) \be
4.4\,\le\,L_{1,A}\,\le\,7.2\,\,\,{\rm fm^3}\, .  \label{EVL} \ee

To assess the global model dependence of the reaction rates for the
breakup processes (\ref{NCN}) and (\ref{CCN}) we consider here the
integral yield \be Y\,=\,\int^\infty_0\,
\Phi_{^{8}B}(E_\nu)\,\sigma(E_\nu)\,dE_\nu\,, \label{Y} \ee where
$\Phi_{^{8}B}(E_\nu)$ is the normalized spectrum corresponding to
the decay of $^{8}B$ \cite{BH} and the cross section $\sigma(E_\nu)$
is given as \be
\sigma(E_\nu)\,=\,\int_{0}^{T_l^{max}}\,\frac{d\sigma}{dT_l}(E_\nu,T_l)\,dT_l\,.
\label{sig} \ee Here $T_l$ is the (kinetic) energy of the outgoing
(charged) lepton. The information on the theoretical uncertainty or
spread of $Y$ is obviously important for the detailed analysis of
the data obtained from the SNO detector.

In Section \ref{prel}, we discuss briefly the methods and inputs
necessary for the calculations and in Section \ref{rd}, we present
the results. We conclude in Section \ref{concl}. Further, in
Appendix \ref{appA}, we present the reaction rates for the charged
channel reaction (\ref{CCN}) with the energy response function of
the SNO detector taken into account, and in Appendix \ref{appB},
we collect the updated cross sections for all deuteron breakup
reactions (\ref{NCN})-(\ref{CCA}) up to (anti)neutrino energies
$E_\nu$=20 MeV.

\section{Methods and inputs}
\label{prel}

To obtain the cross sections one must first calculate
the matrix elements of the weak nuclear currents (charged and neutral)
between the initial and final nuclear states. Here we briefly describe the
needed ingredients of these calculations. We  follow the treatment
described in detail in Section 4 of Ref.\,\cite{MRT1}.

\subsection{Weak nuclear currents}
\label{wnc}

The weak nuclear current used to describe the neutral channel
reaction (\ref{NCN}) is \be j_{NC,\,\mu}\,=\,(1-2
sin^2\theta_W)\,j^3_\mu-2 sin^2\theta_W\, j_{S\mu}+j^3_{5\mu}\,,
\label{WHNC} \ee where $\theta_W$ is the Weinberg angle, $j^3_\mu$
($j^3_{5\mu}$) is the third component of the weak vector (axial)
current in the isospin space, and $j_{S\mu}$ is the isoscalar
vector current. The weak hadron current, triggering the charged
channel reaction (\ref{CCN}), is \be
j^a_{CC,\,\mu}\,=\,j^a_\mu\,+\,j^a_{5\mu}\,,\quad (~a = \pm)~.
\label{WHCC} \ee At low energies, the space component of the weak
axial hadron current is the most important one.

The weak axial nuclear current $j^a_{5\mu}$ for all three
components, $a = \pm $ and 3, consists of the one- and two-nucleon
parts. There is practically no uncertainty associated with the
one-body part. Hence we concentrate on the effects of the two-body
currents. The weak axial nuclear two-body exchange current
$j^a_{5\mu}(2)$ that we consider here is of the OBE type with the
$\pi$-, $\rho$-, $\omega$- and $a_1$ exchanges. It can be divided
\cite{MRT2} into the potential and nonpotential currents. The
potential current of the range $B$, $j^a_{5\mu,B}(2,pot)$, satisfies
the nuclear partially conserved axial current (PCAC) equation, \be
q_\mu j^a_{5\mu,B}(2,pot)\,=\,[V_B,\,j^a_{50}(1)]\,+\,if_\pi
m^2_\pi\Delta^\pi_F(q^2){\cal M}^a_B(2)\,,  \label{WPCAC} \ee where
$V_B$ is the OBEP of the same range $B$, $j^a_{50}(1)$ is the
one-body axial charge density and ${\cal M}^a_B(2)$ is the
associated pion absorption/production exchange amplitude. Further
$f_{\pi}$ is the pion decay constant, $m_{\pi}$ is the pion mass,
and $\Delta_F^{\pi}$ is the pion propagator. This current is model
independent and if a particular OBEP is used to generate the nuclear
wave functions, then its effect can be calculated in a
model-independent way.

The main part of the nonpotential weak axial exchange current
contains the model independent $\rho$-$\pi$ current and the $\Delta$
excitation currents that are model dependent. In our calculations,
we shall adopt the $\pi$-$N$-$\Delta$ and $\rho$-$N$-$\Delta$
Lagrangians used for many years \cite{OO,DMW} to study the $\pi N$
reactions and the pion photo- and electroproduction on a nucleon
(model I) and also the gauge symmetric Lagrangians proposed recently
\cite{PATI,PA} (model II). The $\Delta$ excitation effect is in the
model II suppressed  due to the appearance of an additional factor
$(M/M_\Delta)^2\,\approx\,0.58$ [$M (M_\Delta)$ is the nucleon
($\Delta$ isobar) mass] in the exchange current operators.

Let us note that our model current II differs from an analogous
current of Ref.\,\cite{NSAPMGK}. That current is a purely
phenomenological one, the potential part of which does not satisfy
the PCAC constraint and the suppression of the $\Delta$ strength
is achieved by reducing the $\Delta-N$ coupling to fit the
Gamow-Teller matrix element in the triton beta decay.

\subsection{Nuclear potentials}
\label{np}

We use the Nijmegen I (NijmI), Nijmegen 93 (Nijm93) \cite{SKTS}
and QG \cite {OPT} one-boson-exchange potentials. The couplings
and cutoffs, entering these potentials, are employed in our
exchange currents. For comparison, we also consider the cross
sections calculated from the AV18 potential, which is not an OBEP
(see Table I of Ref.\,\cite{NSGK}).

\subsection{Extraction of $L_{1,A}$}
\label{eloa}

We extract the low energy constant $L_{1,A}$ from comparison of the
cross sections based on the potential models and the EFT form
$\sigma_{EFT}$, see Eq.(\ref{SEFF}). For each of the $i$-th  1-MeV
bins we obtain the $L_{1,A}(i)$ value and take the corresponding
average \be
{L}_{1,\,A}\,=\,\frac{\sum^N_{i=1}\,L_{1,\,A}(i)}{N}\,,\quad
L_{1,\,A}(i)\,=\,\frac{\sigma_{pot,i}\,-\,a_i}{b_i}\,, \label{LB}
\ee where $\sigma_{pot,i}$ is the cross section, calculated in the
potential model and for the {\it i-th} neutrino energy. We use
$N=13$ for the reaction (\ref{NCN}) and $N=14$ for the reaction
(\ref{CCN}), because for the solar neutrinos $E_\nu~\le~15$ MeV and
$i=1$ corresponds to the relevant reaction threshold. In addition,
we extract $L_{1,A}$ also by the least-squares fit. It turns out
that these two values of $L_{1,A}$ are not identical and provide
somewhat different effective cross sections. We label the results
for the reaction rates obtained with $L_{1,A}$ from Eq.\,(\ref{LB})
by $av$, whereas the results calculated with $L_{1,A}$ from the
least square fit will be labeled $lsf$.

\section{Results and discussion}
\label{rd}

To be compatible with the calculations \cite{BCK}, we use the same
weak interaction constants, $G_F=1.166\times 10^{-5}$ GeV$^{-2}$,
$g_A$=-1.26, cos$\theta_C$=0.975, which differ only slightly from
the constants employed in \cite{NSGK}. In Ref.\,\cite{MRT1} the
contributions from the multipole $J=1$ were calculated for the
transition $d~\rightarrow~^{1}S_0$ both for the one- and two-nucleon
currents and also for the transitions
$d~\rightarrow~^{3}P_{j_f},\,j_f=0,1,2$ for the one-nucleon current.
Here the computation code already contains the contributions from
all multipoles $J=0,1,2,3$ and the transitions
$d~\rightarrow~^{2S+1}L_{j_f},\,j_f=0,1,2$ for the one-nucleon
current.

\subsection{Energy dependence}

The extracted values of the low-energy constant $L_{1,A}$ depend on
the way it was determined (averaging or least squares) and on the
potential used. It varied in the limits $3.8 \le L_{1,A} \le 5.7$
fm$^3$ for the neutral current reaction (ncd) (\ref{NCN}) and $3.9
\le L_{1,A} \le 6.4$ fm$^3$ for the charged current reaction (ccd)
(\ref{CCN}). Alternatively, $L_{1,A}$ can be determined by requiring
specifically for the problem of solar $^8$B neutrinos that the
yields $Y$, Eq.\,(\ref{Y}), are identical whether one uses the
corresponding potential model cross section or the EFT one. The
ranges of the $L_{1,A}$ values are then quite similar to those shown
above, namely $4.2 \le L_{1,A} \le 5.6$ fm$^3$ for ncd and $4.4 \le
L_{1,A} \le 6.7$ fm$^3$ for ccd.

In fact, the energy dependent parameter $L_{1,A}(i)$ is not really a
constant (see also Tables 3 and 4 of Ref.\,\cite{MRT1}). Instead its
values varied, even for a fixed choice of the potential and method
of $L_{1A}$ extraction. In other words, this means that the cross
sections evaluated with a single $L_{1A}$, obtained either by
averaging ($av$) or by the least squares fit ($lsf$) as described
above and using  Eq.\,(\ref{SEFF}) differ from the cross sections
based on the potential model in an energy-dependent way. We
illustrate the energy dependence of such differences in Fig.\,1,
where we plot $\delta^a_i$ defined as \be
\delta^a_i\,=\,1-\frac{\sigma_{EFT}[L^{a}_{1,A}(i)]}{\sigma_{pot,i}}\,,
\label{delai} \ee for both methods of  $L_{1,A}$ extraction
($a=lsf,\,av$). For the potential model, we chose the Nijmegen I
potential. Other potential models used in this work provide similar
picture.

\begin{figure}[htb]
\begin{center}
    \leavevmode
     \epsfxsize=.7\textwidth
   \rotatebox{270}{\epsffile{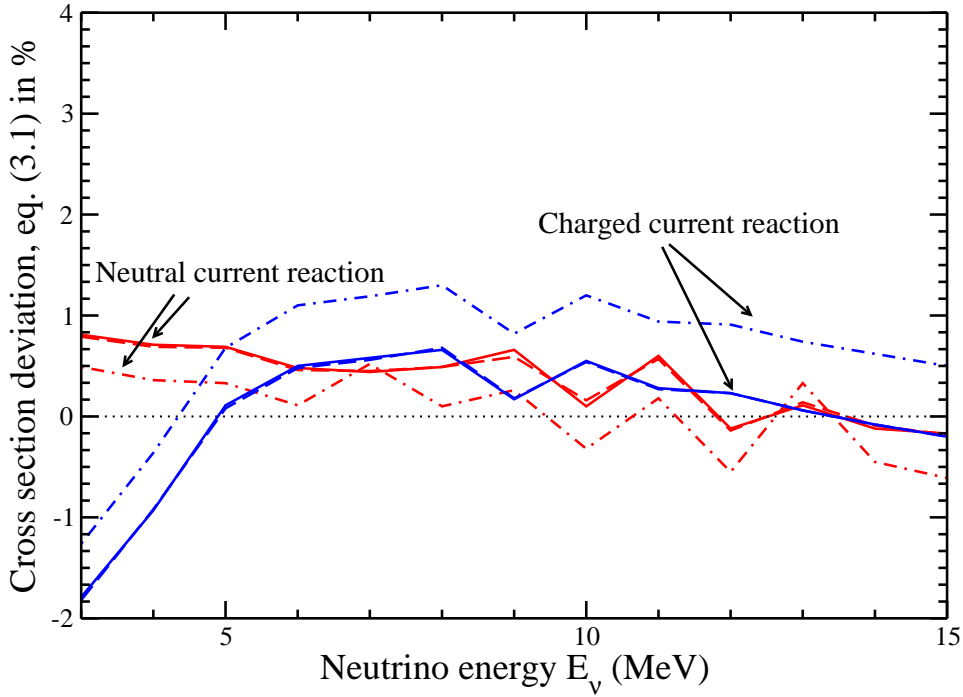}}
\caption{The differences $\delta_i^a$ in percentages [see
Eq.\,(3.1)]. Full lines are based on model I and use $L^{lsf}_{1,A}$
= 5.2 fm$^3$ for ncd and 5.6 fm$^3$ for ccd; the dashed lines are
for the same choice but for the model II, where $L^{lsf}_{1,A}$ =
3.8 fm$^3$ for ncd and 4.2 fm$^3$ for ccd. Finally, the dot-dashed
lines are for the model I with $L^{av}_{1,A}$ = 5.5 fm$^3$ for ncd
and 5.1 fm$^3$ for ccd.}
   \label{fig:1}
\end{center}
\end{figure}

As seen in Fig.\,1 the neutral current cross section behaves in a
regular smooth way, and the EFT and potential model based methods
give cross sections that differ by not more than 1 \% over the
relevant energy range.

The cross section for the charged current reaction exhibits
somewhat stronger variations with energy, in particular for the
lowest energy bins. The origin of this effect remains unknown so
far.

The deviations illustrated in Fig.\,1 cause a corresponding
variations with energy of the partial values $L_{1,A} (i)$. One can
quantify this by pointing out that for the values corresponding to
the full lines in Fig.\,1 the mean-square deviations \be \delta
L_{1,A} = (<L_{1,A}^2> - <L_{1,A}>^2)^{1/2} \ee are $\delta L_{1,A}$
= 0.4, i.e., much smaller than $L_{1,A}^{av}$ = 5.5 for ncd, whereas
for ccd $\delta L_{1,A}$ = 1.8 that is a bit larger but still
considerably smaller than  the $L_{1,A}^{av}$ = 5.1.

\subsection{Global features}

We characterize the global rates by the corresponding reaction
yields (\ref{Y}). The differences between these yields is a
measure of the theoretical uncertainty of the corresponding cross
sections. The results of the calculations for the reactions
(\ref{NCN}) and (\ref{CCN}) are presented in Table
\ref{tab:yield}.

\begin{table}[htb]
\caption{ Reaction rates $Y (\times 10^{-42} {\rm cm}^2)$ for the
weak deuteron breakup  by the  $^8$B neutrinos in the charged (ccd)
and neutral current channels (ncd)  using the model currents I and
II. The yield ratio is $R_i=Y_i(ccd)/Y_i(ncd)$, for i=I, II. In
model II (see Section \ref{wnc}), the $\Delta$ excitation currents
are suppressed by a factor of $\approx$ 0.58. In the columns labeled
by NijmI, Nijm93 and QG, the cross sections are calculated with the
wave functions generated from these potentials (see Section
\ref{np}), the cross sections of the column AV18 are taken from
Table I of Ref.\,\cite{NSGK}. The cross sections of the columns
labeled by $lsf$ and $av$, respectively,  are obtained from
Eq.(\ref{SEFF}) with the constant $L_{1,A}$ calculated by the least
square fit [using Eq.\,(\ref{LB})]. In the last column $\Delta S/S$
is the maximum deviation of the quantity corresponding to the given
row.}
\begin{center}
\begin{tabular}{|l | c || c  c  c | c  c  c |  c  c  c | c  c  c || c |}\hline
     &    & NijmI& $ lsf$&  $av$&Nijm93&$lsf$&$av$ & AV18 &$lsf$&$av$& QG&$lsf$&$av$&$\Delta S/S$ (\%) \\\hline
$Y_I$& ccd & 1.205 & 1.200 & 1.193 & 1.217 & 1.213 & 1.205 & 1.210&1.209&1.207&-&-&-& 1.3\\
$Y_I$&ncd & 0.470 & 0.468 & 0.470 & 0.471 & 0.469 & 0.471 & 0.470 & 0.470 & 0.470&0.470&0.468&0.470 & 0.6\\
$R_I$&    & 2.56  & 2.58  & 2.54  & 2.58  &2.59 &2.56 &2.57&2.57 & 2.57&-&-&-& 2.0\\\hline
$Y_{II}$& ccd & 1.185 & 1.181 & 1.173 & 1.195 & 1.191 & 1.183 &-&-&-&-&-&-& 1.8\\
$Y_{II}$&ncd & 0.462 & 0.460 & 0.462 & 0.462 & 0.460 & 0.462 &-&-&-&0.462&0.460&0.460 &0.4\\
$R_{II}$&    & 2.57  & 2.57  & 2.54  & 2.59 & 2.59 & 2.56&-&-&-&-&-&-&2.0\\\hline
\end{tabular}
\end{center}
\label{tab:yield}
\end{table}

It is seen from Table \ref{tab:yield} that, first of all, despite
the slight energy dependence of the ccd cross section discussed
above, the EFT reaction yields agree reasonably  well with the
corresponding quantities based on the potential models. However,
when $L_{1,A}^{av}$ is used for the ccd channel, the corresponding
yield appears to be systematically smaller. It turns out that the
yield ratios $R_i$ do not depend sensitively on the current model
but depend somewhat more on the choice of the nuclear force and on
the method of the extraction of $L_{1,A}$. In contrast, the
reaction yields depend more on the choice of the current model and
less on the choice of the potential.

As a measure of the uncertainty we shall use the largest relative
difference of the corresponding yield. Using such criterion, an
uncertainty of $\approx$ 2 \% in the calculations of the ratio of
the reaction rates follows, whereas the uncertainty of the reaction
rate is $\approx$ 2.3 \% (3.3 \%) in the neutral (charged) channel
stemming mostly from the difference between the models I and II. Let
us note that the radiative corrections will enhance the reaction
rates by $\approx$ 1.5 \% (2 \%) in the neutral (charged) channel
\cite{KRV}.

In Table \ref{tab:onebody} analogous results obtained when only
the one-nucleon currents are included are presented. It follows
from comparison of the Tables \ref{tab:yield} and Table
\ref{tab:onebody} that the effect of the meson exchange currents,
$\delta^i_{MEC}$, depends on the potential model and varies as
\mbox{$4.6\, \%\,\le\,\delta^I_{MEC}\,\le\,5.7\,\%$} and
\mbox{$2.4\, \%\,\le\,\delta^{II}_{MEC}\,\le\,3.8\,\%$} for the
model currents $I$ and $II$, respectively.

The reaction rates obtained with the one-nucleon currents only still
provide nonvanishing values of $L_{1,A}$ because the strong
interaction part of the problem is treated in the potential model
and in the EFT differently. The corresponding values of the
effective parameter $L_{1,A}$ vary in the limits \be
1.1\,\le\,L_{1,A}\,\le\,2.0\,\,{\rm fm}^3\,. \label{L1AIA} \ee If
one takes in Eq.\,(\ref{SEFF}) $L_{1,A}=0$, then \be
Y^{IA,EFT}_{ccd}=1.120\times 10^{-42}\,{\rm cm}^2, \label{YCC} \ee
for the charged channel reaction, and \be
Y^{IA,EFT}_{ncd}=0.437\times 10^{-42}\,{\rm cm}^2,   \label{YNC} \ee
for the neutral channel reaction. These values are smaller by 2 - 3
\% than the reaction rates of Table \ref{tab:onebody}. However, the
rate ratio $R$ is essentially independent of the two-body current.

\begin{table}[htb]
\caption{ Reaction rates calculated with the one-nucleon currents
only. For the notations see Table \ref{tab:yield}.}
\begin{center}
\begin{tabular}{|l  || c  c  c | c  c  c |  c  c  c  || c |}\hline
        & NijmI& $ lsf$&  $av$&Nijm93&$lsf$&$av$ &  QG&$lsf$&$av$&$\Delta S/S$ (\%) \\\hline
$Y^{IA}_{ccd}$& 1.150 & 1.146 & 1.138 & 1.150 & 1.146 & 1.138 & -&-&-& 1.0\\
$Y^{IA}_{ncd}$& 0.447 & 0.446 & 0.447 & 0.445 & 0.444 & 0.445 & 0.449&0.446&0.449 & 1.1\\
$R$     & 2.57  & 2.57  & 2.55  & 2.58  & 2.58   &2.56  &-&-&-& 1.0\\\hline
\end{tabular}
\end{center}
\label{tab:onebody}
\end{table}

In the charged channel reaction (\ref{CCN}) the electron spectrum
is also measured in SNO. The number of events with the observed
electron kinetic energy $T$ depends then on the response of the
detector function.  The ccd reaction yield is then (see
Ref.\,\cite{SNO4}) \bea
Y_R\,&=&\,\int^\infty_0\,\int^{T_e^{max}}_{0}\,\int^\infty_{T_{th}}\,\Phi_{^8B}(E_\nu)\,\frac{d\sigma}{dT_e}(E_\nu,T_e)\,
R(T_e,T)\,dE_\nu dT_e dT \nonumber \\
\,&\equiv&\,\int^\infty_0\,\Phi_{^8B}(E_\nu)\,\sigma_R(E_\nu)\,dE_\nu\,,
\label{YR} \eea where $T_e$ is the true recoil electron kinetic
energy and $R(T_e,T)$ is the energy response function, \be
R(T_e,T)\,=\,\frac{1}{\sqrt 2\pi
\sigma_T}\,exp\left[-\frac{(T_e-T)^2}{2\sigma^2_T}\right]\,.
\label{ERF} \ee For the pure heavy water phase the resolution width
$\sigma_T$ was taken in the form, \be
\sigma_T(T)\,=\,-0.0648\,+\,0.331\sqrt T \,+\,0.0425 T\,,\quad
T_{th}\,=\,5.0\,MeV\,,  \label{RES} \ee whereas for the salt phase
\cite{SNO2} it was, \be \sigma_T(T)\,=\,-0.145\,+\,0.392\sqrt T
\,+\,0.0353 T\,,\quad T_{th}\,=\,5.5\,MeV\,,  \label{SRES} \ee

To see the effect of the response function and threshold we compare
in Table \ref{tab:spec} the cross section without the response and
the effective cross section $\sigma_R (E_{\nu})$ of Eq.\,(\ref{YR}).
To emphasize the crucial role of the threshold we include a line
corresponding to a hypothetical lower threshold of 5 MeV for the
salt phase.

\begin{table}[htb]
\caption{ The neutrino energy dependent cross sections (in
$10^{-42} {\rm cm}^2$) for the ccd reaction (\ref{CCN}) calculated
using the NijmI potential and the model I currents. The cross
section $\sigma$ is given in Eq.\,(\ref{sig}); $ \sigma_R$-
overlap integral of the cross section with the response function
as defined in the second line of  Eq.\,(\ref{YR}), pure heavy
water phase; $ \sigma^s_R(5.0)$- salt phase with $T_{th}$=5.0 MeV;
$ \sigma^s_R(5.5)$- salt  phase with $T_{th}$=5.5 MeV. The
shorthand $a(-n)$ means $a\times 10^{-n}$.}
\begin{center}
\begin{tabular}{|l | c | c | c| c| c | c | c| c| c | c | c| c| c |}\hline
$E_\nu$ [MeV]& 3 & 4 & 5 & 6 & 7 & 8 & 9 & 10 & 11 & 12 & 13 & 14
& 15  \\\hline $ \sigma $ &  0.0456 & 0.1536 & 0.3406 & 0.6144 &
0.9812 & 1.444 & 2.008 & 2.673 & 3.444 &
            4.322 & 5.310 & 6.410 & 7.622  \\
$ \sigma_R$ & 7.(-7) & 1.(-4) & 0.0051 & 0.0708 & 0.3781 &
            0.9999 & 1.746 & 2.521 & 3.354 & 4.265 & 5.279 & 6.396 & 7.623 \\
$ \sigma^s_R(5.0)$ & 1.(-6) & 2.(-4) & 0.0059 & 0.0749 & 0.3823 &
            0.9984 & 1.743 & 2.518 & 3.352 & 4.263 & 5.277 & 6.393 & 7.619  \\
$ \sigma^s_R(5.5)$ & 4.(-7) & 5.(-5) & 0.0023 & 0.0358 & 0.2361 &
            0.7744 & 1.564 & 2.412 & 3.284 & 4.222 & 5.246 & 6.370 & 7.597 \\\hline
\end{tabular}
\end{center}
\label{tab:spec}
\end{table}

Despite the very important effects of the thresholds and response
function we believe that the global characteristics used in the
Table \ref{tab:yield} can be used as a measure of the theoretical
uncertainty associated with the relative spread of the cross
sections caused be the different model assumptions.

\section{Conclusions}
\label{concl}

We have evaluated the spread of the calculated cross sections, and
of the corresponding reaction yields, for the electron neutrino
from $^8$B decay induced deuteron breakup reactions. The spread is
caused by the different choices of the one-boson-exchange
potentials, and in particular, by the ways the $\Delta$ excitation
currents are treated. Choosing such spread as a measure of the
uncertainty we conclude that the neutral current breakup is $\sim$
2.3 \% uncertain, and the charged current one is $\sim$ 3.3 \%
uncertain. The ratio of the charged to neutral current reaction
rates is then $\sim$ 2 \% uncertain, using this criterion. These
uncertainties are smaller, but basically comparable, to the full
effect of the two-body currents. Thus, we have to conclude that
the evaluation of the effect of the two-body currents remains to
be quite uncertain. We have verified that our  conclusions are not
changed noticeably when the realistic thresholds and resolution
functions of the SNO experiment are used.

\section*{Acknowledgments}
This work was supported by the grant GA \v{C}R 202/06/0746 and by
Ministero dell' Istruzione, dell' Universit\`a e della Ricerca of
Italy (PRIN 2006). Two of us (E.T. and P.V.) thank the INT at the
University of Washington for the hospitality and the Department of
Energy for partial support during the initial stage of this work.
We thank J. Formaggio for very useful discussions.

\newpage

\appendix

\section{Charged channel reaction rates, including the resolution function
and thresholds of the SNO detector }
\label{appA}

Here we repeat some of the previous calculations, but take into
account the resolution function and threshold  of the SNO detector.
The reaction rates for the
charged channel reaction (\ref{CCN}) are presented in Table \ref{tab:YR}.
\par
\begin{table}[htb]
\caption{ The values of the reaction rates $Y_R (\times 10^{-42}$
cm$^{2})$ for the charged current reaction (\ref{CCN}) calculated
according to  Eq.\,(\ref{YR}) and using the potential models NijmI
and Nijm93 and the current models I and II. The values of
$Y^s_{R}$ and $Y^s_{R,IA}$ are calculated with the resolution
function (\ref{SRES}), for comparison the values of $Y^s_{R,IA}$
and $Y_{R,IA}$ are obtained with the one-nucleon currents only.
The ratios R are always obtained by using the total yield from the
column at the left and the related NC total yield either from
Table \ref{tab:yield} or Table \ref{tab:onebody}.}
\begin{center}
\begin{tabular}{|l | c | c || c | c || c | c |}\hline
          & $Y^s_{R,I}$ & $R^s_{R,I}$ & $Y^s_{R,II}$ & $R^s_{R,II}$ & $Y^s_{R,IA}$ & $R^s_{R,IA}$ \\\hline
NijmI  & 0.816 & 1.74 & 0.803 & 1.74 & 0.781 & 1.75  \\
Nijm93 & 0.824 & 1.75 & 0.810 & 1.75 & 0.781 & 1.76
\\\hline\hline
          & $Y_{R,I}$ & $R_{R,I}$ & $Y_{R,II}$ & $R_{R,II}$ & $Y_{R,IA}$ & $R_{R,IA}$ \\\hline
NijmI  & 0.898 & 1.91 & 0.884 & 1.91 & 0.859 & 1.92  \\
Nijm93 & 0.907 & 1.93 & 0.891 & 1.93 & 0.859 & 1.93
\\\hline\hline
\end{tabular}
\end{center}
\label{tab:YR}
\end{table}

It is seen from Table \ref{tab:YR} that the effect of the meson
exchange currents is $4.3\, \%\,\le\,\delta^I_{MEC}\,\le\,6.0\,\%$
and $2.7\, \%\,\le\,\delta^{II}_{MEC}\,\le\,4.0\,\%$ for the model
currents $I$ and $II$, respectively, and it follows closely the
effect obtained above without taking into account the response
function of the detector, though shifted by $\approx$ 0.3 \%
upwards.

The ratio $R=Y_{CC}/Y_{NC}$ was calculated earlier by Bahcall and
Lisi \cite{BL} who obtained \be R\,=\,1.882\,\pm\,0.042\,,
\label{RBL} \ee using the response function (\ref{ERF}) with the
resolution \be \sigma_T(T)\,=\,1.1\sqrt 0.1\, T\,,\quad
T_{th}\,=\,5.0\,MeV\,. \label{RESBL} \ee Adopting such a response
function, we obtained for the NijmI wave functions
$Y_{R,I}(R_{R,I})\,=\,0.889(1.89)$ and
$Y^s_{R,I}(R^s_{R,I})\,=\,0.806(1.71)$, for the pure heavy water
and salt  phases, respectively. The result for $R_{R,I}=1.89$ is
in a very good agreement with  Eq.\,(\ref{RBL}).

Comparison with Table \ref{tab:yield} shows that the reaction yields
$Y^s_{R}$ ($Y_{R}$) are reduced by the factor $\approx$ 0.68 (0.75),
presumably due to the presence of the threshold $T_{th}$. If one
takes $T_{th}=5.0$ MeV for the salt phase, one obtains for
$Y^s_{R,i}$  values that coincide with $Y_{R,i}$ of Table
\ref{tab:YR} within three digits. This is so, because the cross
sections $ \sigma_R$ and $\sigma^s_R(5.0)$ are close to each other
for $E_\nu\,\ge\,7$ MeV (see Table \ref{tab:spec}).

\section{Comparison of the cross sections}
\label{appB}

Here we compare our updated cross sections with the cross sections
of Refs.\,\cite{NSGK} and \cite{BCK} up to (anti)neutrino energies
$E_\nu$=20 MeV. Tables \ref{tab:3} and \ref{tab:4} supersede
Tables 3 and  4 of Ref.\,\cite{MRT1} and Tables 3 - 6 of
Ref.\,\cite{MRT3}.
\par
\begin{table}[htb]
\caption{Cross sections and the differences, in percentages, between
the cross sections for the reactions (\ref{NCN}) and (\ref{NCA}). In
the first column, $E_\nu$ [MeV] is the neutrino energy, in the
second column, $\sigma_{NijmI}$ (in $10^{-42}\times$ cm$^2$) is the
cross section,  calculated with the NijmI nuclear wave functions,
$g_A$=-1.26 and $G_F=1.166\times 10^{-5}\,{\rm GeV}^{-2}$, i.e., the
weak interaction parameters used in Ref.\,\cite{BCK}. In column 3 is
the difference between $\sigma_{Nijm I}$ (I) and the EFT cross
section (\ref{SEFF}) $\sigma_{EFT}$, calculated with the
corresponding constant $L^{av}_{1,\,A}$ given in the parentheses.
The difference between  $\sigma_{NSGK}$ taken from Table I of
Ref.\,\cite{NSGK} and $\sigma_{EFT}$ is given in column 4 (N).
Further, $\Delta_{1(2)}$ is the difference  between the cross
sections $\sigma_{NijmI}$ ($\sigma_{Nijm93}$) and $\sigma_{NSGK}$.
In this case, our cross sections are calculated with $g_A$=-1.254
\cite{NSGK}. The second part of the table  is an analogue  for the
reaction (\ref{NCA}).}
\begin{center}
\begin{tabular}{|l c c c c c ||l c c c c c |}\hline
  & $ \nu_x$  +  $d$ & $\longrightarrow $ & ${\nu_x}'$ +  $n\,p$ & &
& &  ${\bar \nu}_x$  +  $d$ & $\longrightarrow $ & ${{\bar
\nu}_x}'$ +  $n\,p$ & &
\\\hline
$E_\nu$ & $\sigma_{NijmI}$&I (5.3)&N (5.4)
&$\Delta_{\,1}$&$\Delta_{\,2}$& $E_{\bar{\nu}}$&$\sigma_{NijmI}$&I
(5.6)&N (5.5)&$\Delta_{\,1}$&$\Delta_{\,2}$
\\\hline\hline
3&0.00335&0.6&0.4&-0.9&-0.5& 3&0.00332&0.0&0.1&-0.9&-0.4
\\
4&0.0307 &0.6&0.2&-0.6&0.3& 4&0.0302 &0.5&0.2&-0.5&0.4
 \\
5&0.0949 &0.5&0.2&-0.8&-0.4& 5&0.0930 &0.3&0.1&-0.7&-0.3
 \\
6&0.201  &0.3&0.1&-0.8&-0.9& 6&0.196  &0.5&0.3&-0.7&-0.8
 \\
7&0.353  &0.3&0.1&-0.9&-1.0& 7&0.343  &0.1&0.1&-0.8&-0.9
 \\
8&0.553  &0.3&0.2&-0.9&-0.6& 8&0.533 &0.9&0.8&-0.7&-0.3
 \\
9&0.801  &0.5&0.4&-1.0&-0.8& 9&0.768 &0.3&0.2&-0.8&-0.5
 \\
10&1.099 &-0.1&-0.1&-1.0&-0.9& 10&1.049&0.2&0.2&-0.8&-0.7
 \\
11&1.447 &0.4&0.5&-1.1&-1.0& 11&1.373&-0.2&-0.2&-0.9&-0.8
 \\
12&1.848&-0.3&-0.3&-1.1&-0.7& 12&1.744&-0.3&-0.4&-0.8&-0.4
 \\
13&2.299&-0.1&0.0&-1.2&-0.9& 13&2.158&-0.3&-0.2&-0.9&-0.6
\\
14&2.802&-0.2&0.0&-1.3&-1.0& 14&2.616&-0.2&-0.2&-0.9&-0.6
 \\
15&3.359&-0.3&-0.1&-1.3&-1.1& 15&3.118&-0.3&-0.2&-1.0&-0.7
 \\
16&3.968&-0.5&-0.3&-1.4&-1.2& 16&3.663&-0.2&-0.1&-1.0&-0.8
\\
17&4.631&-0.7&-0.4&-1.4&-1.3& 17&4.252&-0.4&-0.2&-1.0&-0.9
 \\
18&5.348&-0.6&-0.3&-1.4&-1.6& 18&4.882&-0.4&-0.3&-1.1&-1.2
 \\
19&6.119&-0.7&-0.4&-1.5&-1.6& 19&5.555&-0.5&-0.3&-1.1&-1.3
 \\
20&6.949&-0.9&-0.6&-1.5&-1.7& 20&6.273&-0.5&-0.2&-1.2&-1.3
 \\\hline
\end{tabular}
\end{center}
\label{tab:3}
\end{table}
\newpage
\par
\begin{table}
\caption{ Cross sections and the differences, in percentages,
between the cross sections for the reactions (\ref{CCN}) and
(\ref{CCA}). For the notation, see Table \ref{tab:3}. In addition,
cos$\theta_C$=0.975 is used for comparison with the EFT, whereas
cos$\theta_C$=0.9749 when we compare our cross sections with
Ref.\,\cite{NSGK}.}
\begin{center}
\begin{tabular}{|l c c c c c ||l c c c c c |}\hline
 & $ \nu_e$  +  $d$ & $\longrightarrow $ & $e^-$ +  $p\,p$ & &
& & ${\bar \nu}_e$  +  $d$ & $\longrightarrow $ & $e^+$ +  $n\,n$
& &
\\\hline
$E_\nu$ &$\sigma_{NijmI}$&I (5.1)&N
(6.0)&$\Delta_{\,1}$&$\Delta_{\,2}$&
$E_{\bar{\nu}}$&$\sigma_{NijmI}$&I (5.2)&N
(5.6)&$\Delta_{\,1}$&$\Delta_{\,2}$
\\\hline\hline
2&0.00341&-5.5&-0.7&-6.7&-5.9&
2&-&-&-&-&- \\
3&0.0456 &-1.2&-0.4&-2.7&-1.9&
3&-&-&-&-&- \\
4&0.154  &-0.4&-0.6&-1.7&-0.8&
4&-&-&-&-&- \\
5&0.341  & 0.6& 0.1&-1.4&-0.5&
5&0.0274 &-2.0&-1.0 &-2.3&-1.5 \\
6&0.614  & 1.0& 0.3&-1.3&-0.4&
6&0.117  &-0.6 &-0.1 &-1.9&-1.1 \\
7&0.981  & 1.1& 0.4&-1.3&-0.3&
7&0.278  &-0.4 &-0.2 &-1.6&-0.7 \\
8&1.444  & 1.2& 0.5&-1.3&-0.3&
8&0.515  &-0.1 &-0.1 &-1.4&-0.4 \\
9&2.008  & 0.7& 0.0&-1.4&-0.3&
9&0.832  &-0.1 &-0.2 &-1.3&-0.3 \\
10&2.673 & 1.1& 0.5&-1.4&-0.4&
10&1.230 &0.5  &0.3 &-1.3&-0.2 \\
11&3.444 & 0.8& 0.3&-1.5&-0.5&
11&1.708 &0.4  &0.2 &-1.3&-0.2 \\
12&4.322 & 0.9& 0.3&-1.6&-0.5&
12&2.265 &0.4  &0.1 &-1.2&-0.1 \\
13&5.310 & 0.7& 0.2&-1.6&-0.6&
13&2.903 &0.3  &0.0 &-1.2&-0.1 \\
14&6.410 & 0.6& 0.2&-1.7&-0.6&
14&3.618 &0.5  &0.2 &-1.2&-0.1 \\
15&7.622 & 0.5& 0.1&-1.7&-0.6&
15&4.411 &0.3  &0.0 &-1.3&-0.1 \\
16&8.936 &0.1 &-0.1&-1.9&-0.8&
16&5.280& 0.3  &0.1 &-1.3&-0.1 \\
17&10.37 & 0.0&-0.2&-2.1&-1.0&
17&6.225& 0.4  &0.2 &-1.3&-0.2 \\
18&11.93 &-0.1&-0.1&-2.1&-1.1&
18&7.244& 0.5  &0.4 &-1.4&-0.2 \\
19&13.61 &-0.1&-0.1&-2.2&-1.1&
19&8.335& 0.3  &0.2 &-1.4&-0.2 \\
20&15.42 &-0.2&-0.3&-2.2&-1.1& 20&9.498& 0.4  &0.3&-1.5&-0.3
\\\hline
\end{tabular}
\end{center}
\label{tab:4}
\end{table}

\end{document}